\documentclass[%
 reprint,
 amsmath,amssymb,
 aps,
]{revtex4-2}
\usepackage[normalem]{ulem}
\usepackage{xcolor}
\usepackage{graphicx}
\usepackage{dcolumn}
\usepackage{bm}
\usepackage{hyperref}
\usepackage{soul}

\newcommand{\edits}[1]{\textcolor{black}{{#1}}}
\newcommand{\ed}[1]{\textcolor{black}{{#1}}}

\usepackage{lineno}

\begin{document}

\preprint{APS/123-QED}


\title{\edits{
Beyond Dipolar Activity: Quadrupolar Stress Drives Collapse of Nematic Order on Frictional Substrates}
}

\author{Aleksandra Arda\v{s}eva$^1$}
 \thanks{These authors contributed equally to this work}
\author{Ignasi V\'{e}lez-Cer\'{o}n$^{2,3}$}%
 \thanks{These authors contributed equally to this work}
\author{Martin Cramer Pedersen$^1$}%
\author{Jordi Ign\'{e}s-Mullol$^{2,3}$}%
\email{jignes@ub.edu}
\author{Francesc Sagu\'{e}s$^{2,3}$}%
\email{f.sagues@ub.edu}
\author{Amin Doostmohammadi$^1$}%
\email{doostmohammadi@nbi.ku.dk}
\affiliation{$^1$Niels Bohr Institute, University of Copenhagen, Blegdamsvej 17, Copenhagen, Denmark\\$^2$Department of Materials Science and Physical Chemistry, Universitat de Barcelona, Barcelona 08028, Spain\\$^3$Institute of Nanoscience and Nanotechnology, IN2UB, Universitat de Barcelona, Barcelona 08028, Spain
}%


\begin{abstract}
\edits{The field of active nematics has traditionally employed descriptions based on dipolar activity, with interactions that align along a single axis. However, it has been theoretically predicted that interactions with a substrate, prevalent in most biological systems, would require novel forms of activity, such as quadrupolar activity, that are governed by hydrodynamic screening. Here, by combining experiments and numerical simulations, we show that upon light-induced solidification of the underlying medium, microtubule-kinesin mixtures undergo a transformation that leads to a biphasic active suspension. Using an active lyotropic model, we prove that the transition is governed by screening effects that alter the dominant form of active stress. Specifically, the combined effect of friction and quadrupolar activity leads to a hierarchical folding that follows the intrinsic bend instability of the active nematic layer.}
Our results demonstrate the dynamics of the collapse of orientational order in active nematics and present a new route for controlling active matter by modifying the activity through changing the surrounding environment.
\end{abstract}

\maketitle

Living materials, such as eukaryotic cells and bacteria, are distinct from their inanimate counterparts in their ability to continuously convert chemical energy to mechanical work at the level of their constituent elements~\cite{prost2015active}. This {\it activity}, in turn, orchestrates the formation of collective patterns of motion on scales much larger than the size of the individuals~\cite{hartmann2019emergence, needleman2017active, doostmohammadi2018active,sagues23}, and guides important physiological processes, such as organ development~\cite{bianco2007two, maroudas2021topological} and collective invasion~\cite{ariel2015swarming, clark2015modes, zhang2022topological}. Similar activity-induced collective patterns are increasingly emulated in bio-inspired synthetic materials, with the aim of rendering them capable of self-organization and self-healing~\cite{sanchez2012spontaneous, schaller2010polar}.

In the bulk, the mechanistic basis and fundamental instabilities of activity-induced collective pattern formation have been intensely studied both in theoretical models~\cite{ramaswamy2010mechanics, marchetti2013hydrodynamics, alert2022active} and in experiments~\cite{doostmohammadi2022physics}. Importantly, however, living materials in most natural setups are constrained by confining boundaries and substrates, so that the dynamics of the active systems are closely controlled through their interaction with the surrounding environment. Striking examples include subcellular active flows confined to the cell boundaries~\cite{callan2016actin}, twitching motility of surface-attached bacteria~\cite{gomez2023substrate}, and dynamics of cellular monolayers interacting with their underlying substrate~\cite{krishnan2011substrate}.

The dynamics of active matter confined to channels of varying shapes and sizes have been the focus of intense research. While various modes of collective self-organization have been identified in theoretical models~\cite{doostmohammadi2017onset, shendruk2017dancing, samui2021flow, coelho2019active} and experimental systems including subcellular filaments~\cite{hardouin2019reconfigurable, opathalage2019self, hardouin2022active}, bacteria~\cite{lushi2014fluid, you2021confinement}, and confined epithelial and mesenchymal cells~\cite{deforet2014emergence, duclos2018spontaneous}, the impact of active matter interaction with a tunable and dynamically changing environment is not well understood. 
Theoretical studies, focusing on the frictional and momentum-damping effects induced by substrates, have suggested possibilities of stabilization, control, and significant alterations in steady-state collective patterns of active materials~\cite{thampi2014active, thijssen2020role, ramaswamy2003active, thijssen2021submersed,zhang2021spatiotemporal}. Nevertheless, experimental realization of such effects and deciphering the complex impact of the substrate beyond frictional effects, has remained a significant challenge. Moreover, while the focus so far has been on steady-state patterns of motion, the study of transient effects of active matter interacting with substrates, and the possibility of transitioning between different steady-states are yet to be explored.

Here, using a mixture of microtubule-kinesin motor proteins as a model active system residing on a tunable hydrogel material, we reveal a novel hierarchical transition that extinguishes the nematic order of the system and transforms it into a biphasic, yet still active, suspension of the bundled microtubules. 
We combine experimental observations with a continuum model of lyotropic active nematic to reveal the mechanisms that govern the transition between these two dynamical steady states and uncover the previously overlooked impact of the substrate on tuning the dominant forms of active stresses.

\begin{figure*}[htb!]
\centering
\includegraphics[width=1\linewidth]{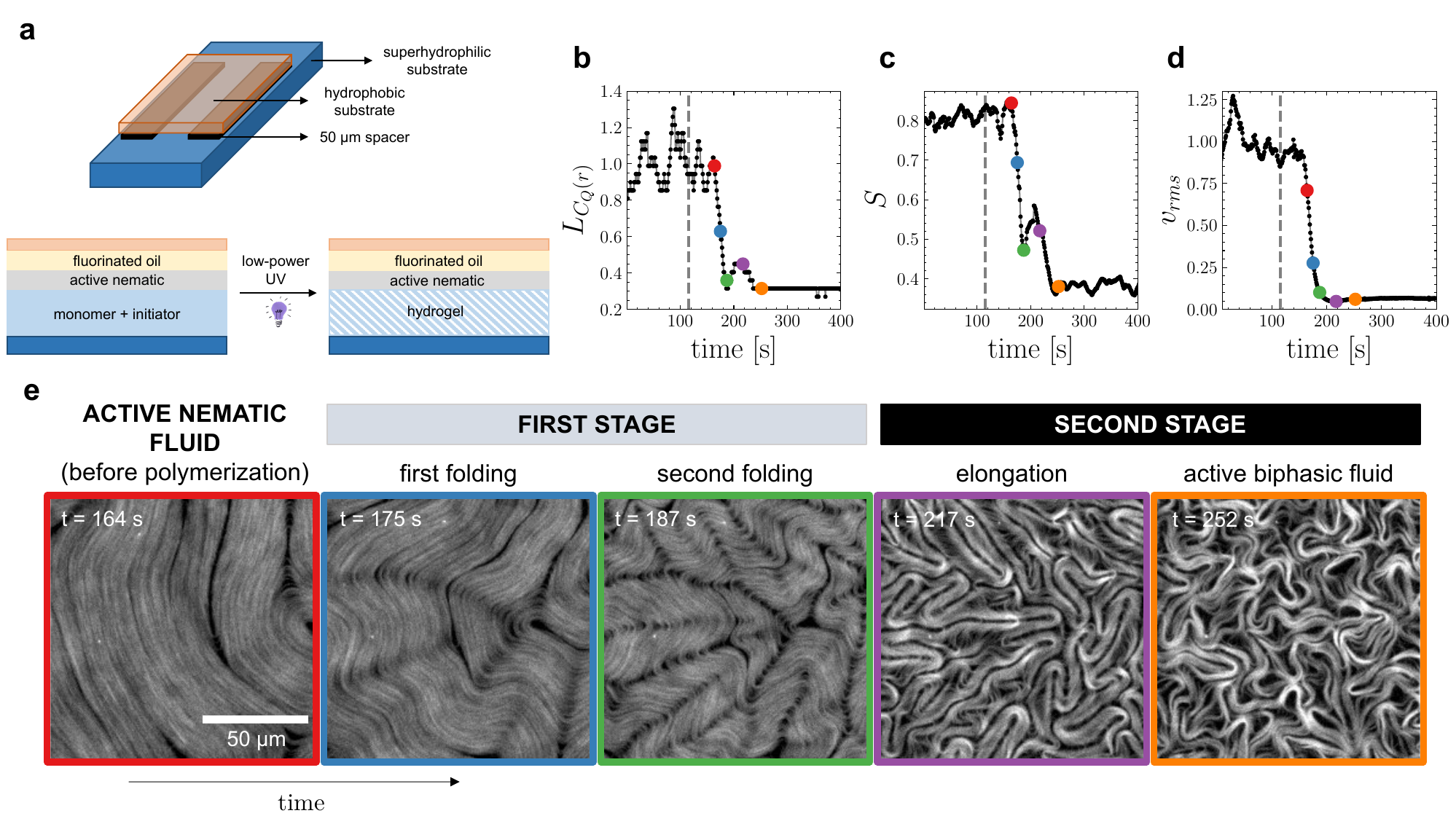}
\caption{\label{fig1} \textbf{Activation of light leads to a two-stage dynamic transition.} (a) Schematic representation of the experimental cell. The flow cell is filled with fluorinated oil and an aqueous solution containing the active material, monomer, and initiator of the hydrogel. A flow cell of 50 $\mu$m thickness is prepared using a superhydrophilic slide and a hydrophobic cover slide. Low-power UV light leads to the formation of the hydrogel, confining the active material between it and the oil phase. \edits{In this experiment, [ATP]=1500 $\mu$M, and light power density is 3 mW cm$^{-2}$.} (b) -- (d) Director correlation length, $L_{C_Q}(r)$, order parameter, $S$, and root-mean-squared velocity, $v_{\text{rms}}$, respectively, as a function of time. The gray dashed line marks the time when the light is turned on. $L_{C_Q}(r)$ and $v_{\text{rms}}$ are scaled by pre-light averages. (e) Snapshots from experiments, showing patterns for different transition checkpoints. The color of each frame corresponds to the corresponding time point indicated in panels (b)--(d) (see also Movie S1).}
\end{figure*}

In order to explore the effect of contact with the substrate in a controllable manner, we designed an experimental system where we employ the standard formulation of a kinesin/tubulin active gel to which we add the precursors of a poly(ethylene glycol) (PEG)-based hydrogel that can be photopolymerized with UV light with the concourse of a photoinitiator \cite{Velez24}\edits{. We have verified that the power of the used UV radiation has no direct effect on the morphology or dynamics of the AN layer}. Experiments are performed in a flow cell of 50 µm thickness, made with a superhydrophilic and a hydrophobic glass slide (Fig. \ref{fig1}a). The cell is filled with the active mixture and fluorinated oil to allow the formation of the active nematic at the water-oil interface. The interfaced material displays, at this stage, the well-established, seemingly chaotic flow patterns characteristic of an active nematic layer~\cite{sanchez2012spontaneous,guillamat2016control}. We then subject the cell to a sustained (120 s) low-power-illumination. 
The hydrogel polymerizes, leaving a wetting film that is still capable of accommodating the microtubule-based fluid. \edits{In our experiments, we changed activity by controlling the concentration of ATP, which we kept above $300\,\mu M$ to prevent the arrest of the active nematic upon substrate polymerization. \ed{See \cite{supplementary} for further details on the experimental protocols}}. 
\begin{figure*}[htb!]
\centering
\includegraphics[width=\linewidth]{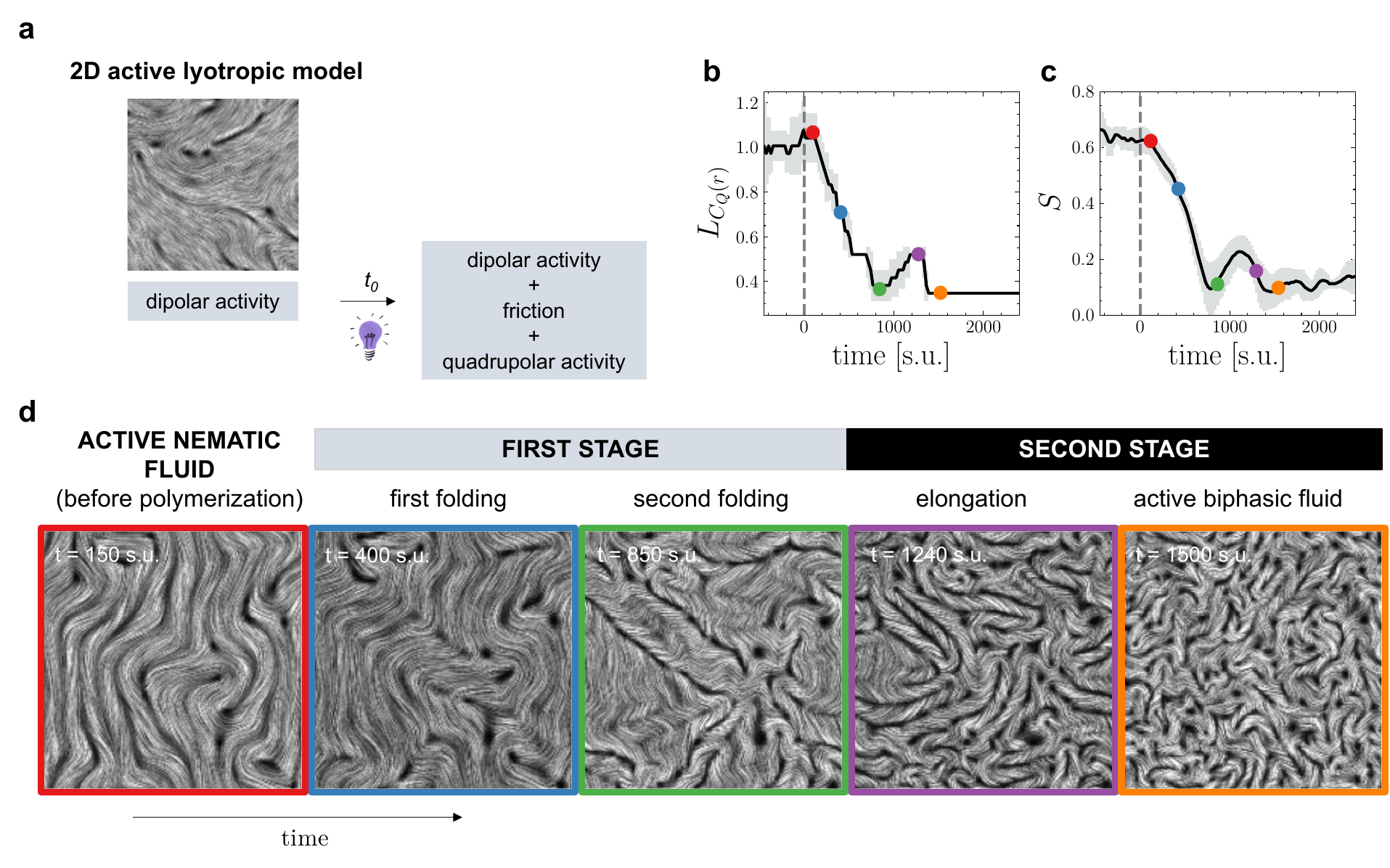}
\caption{\label{fig2} \textbf{Continuum simulations reproduce the two-stage transition from active nematic fluid to active biphasic fluid.} (a) Schematic depiction of the simulation setup. In the beginning, simulation is carried out with only dipolar activity. At $t_0$, in addition to the dipolar activity, friction and quadrupolar activity are turned on to mimic the effect of turning on light in the experiments. (b)--(c) Director correlation length (rescaled by the value before $t_0$), $L_{C_Q (r)}$, and the order parameter, $S$, respectively, as a function of time (in simulation units, s.u.). The gray dashed line denotes the onset of friction and quadrupolar activity ($t_0^f=t_0^q$). Plots are averaged over 5 realizations. (d) Line integral convolution (LIC) render of snapshots of the director at different stages of transition (see Movie S2). Black regions correspond to the isotropic phase. The color of each frame corresponds to the corresponding stage indicated in panel (b). \ed{The parameters used are listed in \cite{supplementary}.}}
\end{figure*}

The observed changes in the orientational textures following this simple intervention are drastic (see Fig. \ref{fig1}e and Supplementary Movie S1). The material starts to fold in on itself, following what can be interpreted as a cascade of hierarchical bending instabilities. The result is a highly corrugated material, as exhibited by the practical collapse of its orientation two-point correlations (see Fig. \ref{fig1}b). In the final state, the microtubule suspension seems to rarefy while losing nematic order (see Fig. \ref{fig1}c). We term this state {\em active biphasic fluid}, as despite being largely parceled by isotropic non-fluorescent domains, the material does not cease to move at any time, but at much lower velocities (see Fig. \ref{fig1}d). 
A closer inspection of the temporal evolution of the system reveals that upon turning on the light the crossover from active nematic to active biphasic fluid goes through a two-stage process: a cascade of self-similar folding patterns, followed by a drop in the orientational order and increase in the number of isotropic domains that span the system (see Supplementary Fig. S2a). A hallmark of this two-stage transition is also evident in the evolution of the director correlation length with time (see Fig. \ref{fig1}b): as hierarchical folding emerges during the first stage of the transition, the correlation length and the magnitude of the orientational order sharply drop (Fig. \ref{fig1}c). This is followed by a subsequent increase and a peak in the correlation length, as the isotropic domains start to elongate, which is then followed by a second drop in the correlation length and the magnitude of the orientational order. \edits{Additionally, an increase in the defect density and a reduction in the size of active flows around topological defects is observed (Supplementary Fig. S3)}.
This identifies the second stage of the transition leading to an active biphasic fluid state being established. 

To investigate the physical mechanism driving the experimentally observed transition, we used a continuum model of lyotropic active nematics~\cite{blow2014biphasic}. Here, a non-conserved scalar parameter, $\phi \in [0,1]$, is introduced to distinguish the active nematic phase driven by dipolar active stresses~\cite{ramaswamy2003active,marenduzzo2007steady,martinez2019selection}, from a passive, isotropic phase. 
When the UV light is on, the gel underneath the substrate solidifies. This leads to a friction-induced screening effect that affects the dominant forms of active stresses exerted by active entities~\cite{mathijssen2016hydrodynamics}. Therefore, higher-order contributions to active stresses become important~\cite{mathijssen2016hydrodynamics,maitra2018nonequilibrium,sultan2022quadrupolar}. To account for this, we introduce quadrupolar activity to the momentum equation, together with the isotropic friction \ed{\cite{supplementary}}. To replicate the experimental condition as closely as possible, these terms are switched on instantaneously at a given time after the system has reached the active turbulence state (Fig \ref{fig2}a). This corresponds to switching the UV light on in the experiment.
\begin{figure*}[htb!]
\centering
\includegraphics[width=\linewidth]{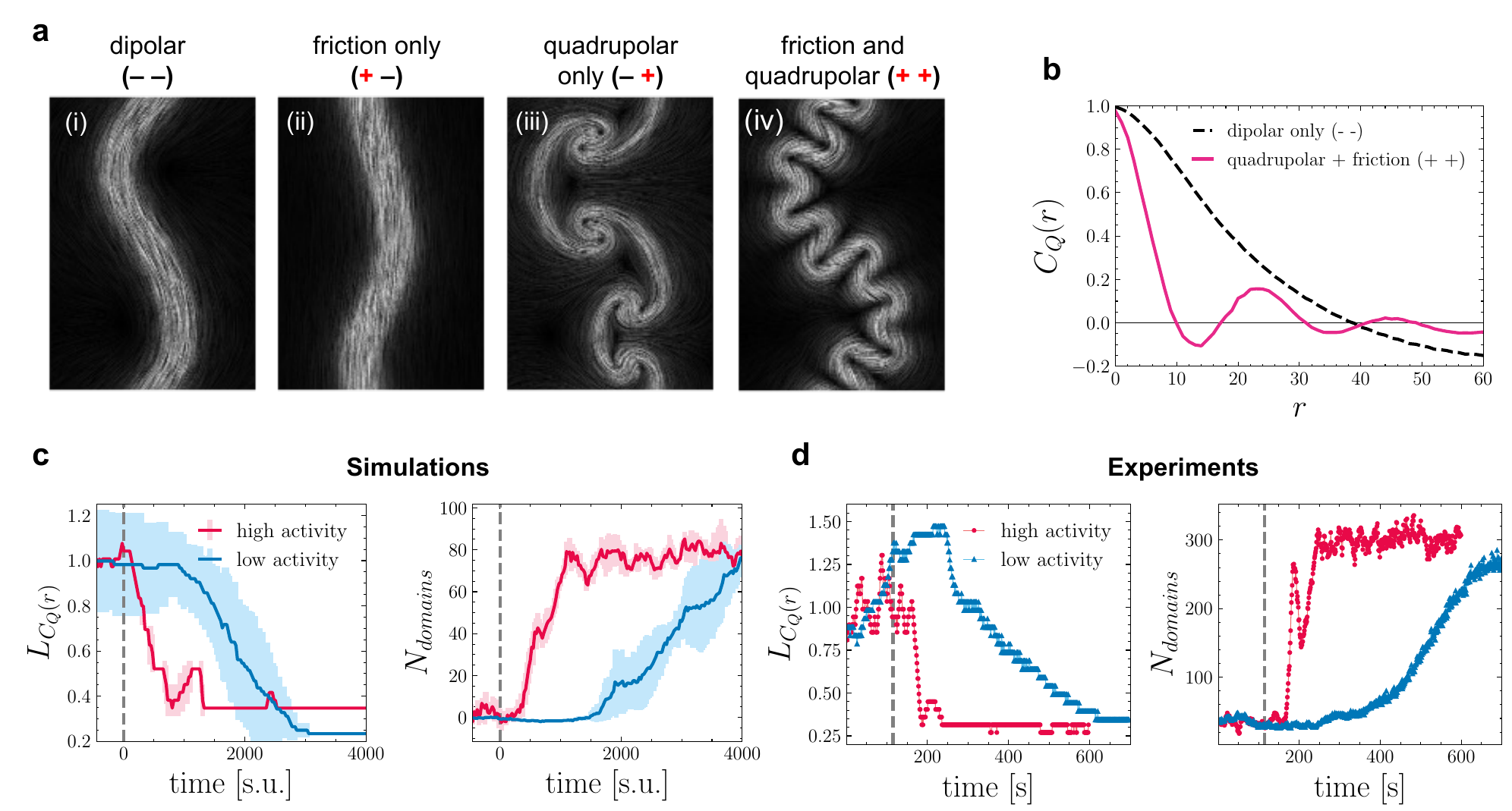}
\caption{\label{fig3} \textbf{\edits{Quadrupolar activity governs the} physical mechanism of the dynamic transition.} (a) LIC render of a \edits{simulated} stripe of nematic phase surrounded by isotropic phase (black) with bend instability evolved under different conditions: (-- --) no friction and no quadrupolar activity; (+ --) friction only; (-- +) quadrupolar activity only; (+ +) both friction and quadrupolar activity. Dipolar activity is on at all times. See Movies S3--S6. (b) Correlation of director before the friction and quadrupolar activity are activated (black dashed line) and after (pink line). (c) Director correlation length (rescaled by pre-light average) and the order parameter as a function of time for high and low values of dipolar activity obtained via simulations. Plots are the average over 5 realizations. \edits{Here, $\zeta_{dipole}=\zeta_{dipole}^0$ (high activity) and $\zeta_{dipole}=0.1\zeta_{dipole}^0$ (low activity).} \ed{The parameters used are listed in \cite{supplementary}.} (d) Same as (c) but for experiments (single realization, see Movie S7). \edits{Light power density is 3 mW cm$^{-2}$, and [ATP]=1500 $\mu$M for high activity and 300 $\mu$M for low activity conditions.}%
}
\end{figure*}

The model qualitatively reproduces the dynamics in the director field (see Fig. \ref{fig2}d and Supplementary Movie S2). Ordered regions undergo folding, breaking into smaller perpendicular domains. Afterwards, the domains elongate until the active biphasic regime with smaller length scales is established. Remarkably, not only do the qualitative features mirror experimental observations, but the transition follows a two-stage dynamics as both the two-point director correlation and the order collapse (Fig. \ref{fig2}b--c), and the number of isotropic domains subsequently increases in agreement with experiments (Supplementary Fig. S2b). Additionally, in line with the experimental measurements, the crossover between the two stages is marked by the temporal, step-like (see Supplementary Fig. S4) increase and a peak in the director correlation length before the transition to biphasic active fluid, \edits{as well as decrease in the root-mean-squared velocity (Supplementary Fig. S5).}
Together, the experimental observations and continuum modeling reveal a novel two-stage process for the dynamic transition between active nematic and active biphasic fluids that is mediated by the interaction with the substrate.

Next, we investigate the mechanism for the emergence of self-similar patterns. To understand the separate roles of friction and quadrupolar activity in inducing hierarchical folding, we simulate a minimal setup that consists in setting a stripe of low-activity nematic fluid surrounded by an isotropic phase (Supplementary Movies S3--S6). Here, we let the system evolve through the well-established bend instability (Fig. \ref{fig3}a, panel (i)) under dipolar activity only, and then turn on friction and/or quadrupolar activity. 

Turning on only isotropic friction stabilizes the system and suppresses the growth of the bend instability (Fig. \ref{fig3}a, panel (ii)). On the other hand, when quadrupolar activity acts on the system (and the friction is turned off), the nematic phase folds in on itself, creating mushroom-like structures (Fig. \ref{fig3}a, panel (iii)). 
Finally, when both friction and quadrupolar activities are turned on (Fig. \ref{fig3}a, panel (iv)), the hierarchical buckling of the existing bend instability is observed. This secondary buckling occurs at a much smaller wavelength, compared to the primary bend instability, as shown by the measurements of the orientation correlation (see Fig. \ref{fig3}b). These results indicate that the observed patterns during the two-stage transition result from a balance between the dipolar and quadrupolar activities (destabilizing) and friction (stabilizing) that leads to cascades and eventual decreases in length scales. Hence, the ratios between these three parameters determine the possibility of a substrate-induced transition. \edits{(see Supplementary Fig. S1 for a detailed parameter study)}.

\begin{figure}[htb!]
\centering
\includegraphics[width=0.9\linewidth]{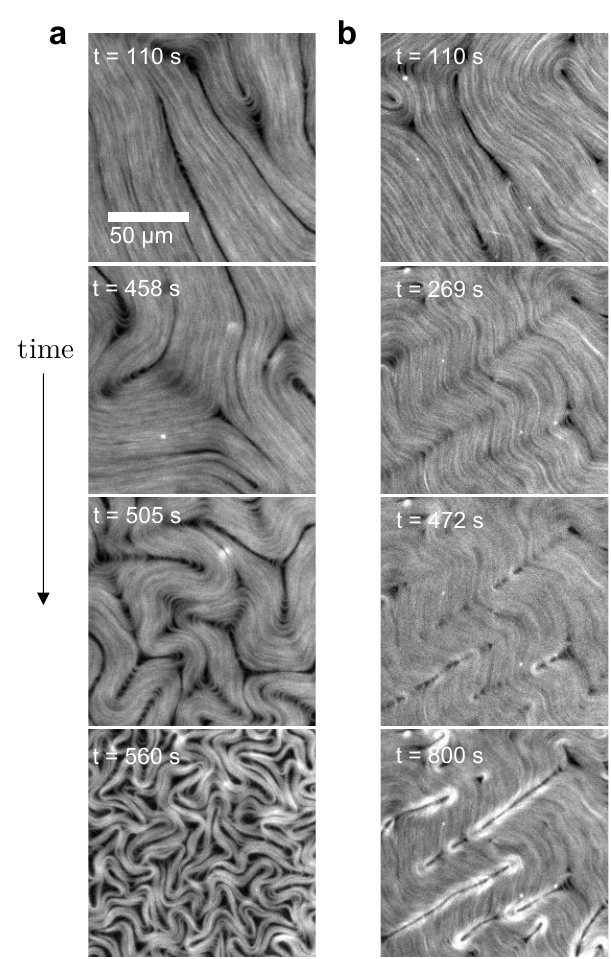}
\caption{\label{fig4} \edits{\textbf{Tuning the transition via UV light power.} Snapshots from experiments performed at a lower light power density of 0.3 mW cm$^{-2}$, showing patterns for different transition checkpoints for (a) high activity ([ATP]=1500 $\mu$M) and (b) low activity ([ATP]=300 $\mu$M) regimes.}}
\label{fig:low}
\end{figure}

Having established the mechanism of the two-stage collapse of the orientation order in active nematics, we use the model to explore the possibility of controlling the dynamics by varying the activity of the system. 
The model (Fig. \ref{fig3}c) predicts that for a less active system with lower dipolar activity, the two-stage transition \edits{is not observed, and} the decay to smaller length scales and collapse of the order are slowed down. The increase in the number of isotropic domains is also slowed down (see Supplementary Fig. S2b). Similar behavior can be seen in the experiments (Fig. \ref{fig3}d) when the activity of the microtubules is reduced by lowering ATP concentration before shining light. The resulting experimental patterns (see Supplementary Movie S7) show similar stages of the transition, including folding and elongation. The temporal evolution of the correlation length, order parameter, and number of domains qualitatively confirm the model predictions (Fig. \ref{fig3}d and Supplementary Fig. S2a). Together, the modeling and experimental results show how the interplay of dipolar activity with frictional damping and quadrupolar activity shapes the dynamics of the transition from active nematic to active biphasic fluid.

\edits{Finally, we explored the impact of UV light power on the transition. When the activity is high (Fig. \ref{fig:low}a, Supplementary Movie 8), the transition from active nematic to active biphasic fluid is modified and the self-folding cascade is not observed. However, when the activity is low (Fig.~\ref{fig:low}b, Supplementary Movie 9), the system freezes while defects continue moving slowly. The accumulation of microtubules around defect cores is observed. In the simulations, these behaviors, including the accumulation of nematic fluid around isotropic elongated domains, are reproduced by varying the ratio of friction, $f$, to quadrupolar activity, $\zeta_{quad.}$ (see Supplementary Fig. S6). 
These qualitative changes in the behavior of the system upon lowering the light intensity can be best understood by considering the competition between the time scales set by dipolar and quadrupolar activity, i.e. $\tau_\text{dipole} \sim \eta / \zeta_\text{dipole}$ and $\tau_\text{quad.} \sim \eta / \zeta_\text{quad.}$, respectively, with the time scale set by friction $\tau_f \sim \rho / f$. At the same level of high dipolar activity, reducing friction and quadrupolar activity results in a dynamic that is primarily set by the dipolar activity time scale, and as such the self-folding cascade is not observed. At the same low light level, lowering the dipolar activity can result in the frictional time scale dominating the dynamics and arresting the motion, while at this low light intensities, the quadrupolar activity remains low enough that it cannot induce the self-folding cascades.}

Our continuum model of lyotropic active fluid reveals the mechanistic basis of the evolution of the system upon solidification of the substrate and shows that increased friction alone is insufficient to explain the observed transition. Substrate friction not only extracts momentum from the system but also induces screening effects, which alter the dominant form of active stress from dipolar to quadrupolar stress. The emergence of significant quadrupolar contribution, induced by hydrodynamic screening effect of the substrate, is in line with theoretical predictions in the context of single active particles~\cite{mathijssen2016hydrodynamics} and dense collections of active particles, both in the presence of hydrodynamics~\cite{maitra2018nonequilibrium} and in over-damped systems~\cite{sultan2022quadrupolar}. To our knowledge, the experiments presented in this work provide the first evidence of the important 
role of quadrupolar active stresses in dense active matter systems under friction. The continuum model further demonstrates how such quadrupolar activity combined with frictional damping can induce the folding of active domains onto themselves, leading to the emergence of cascades of self-similar folding patterns. 

Our findings present \edits{the first example demonstrating that current models for active nematic based on dipolar activity are not enough to account for the changes that ensue a modification of the substrate mechanical properties.}
\edits{Unlike other studies, where shining the light modified the properties of active nematic itself \cite{lemma2023spatio, zarei2023light, zhang2021spatiotemporal}, our experiments allowed to control only the physical properties of the substrate.}
Understanding the dynamics of active materials, such as cells and bacteria, requires accounting for the dynamic changes in their surrounding environment. 
Examples like the extracellular matrix (ECM) and bacterial biofilms underscore the importance of this dynamic interplay. The ECM, with its ability to transition between fluid-like and solid-like states, profoundly influences cell behavior, tissue remodeling, and disease progression~\cite{elosegui2021extracellular}. Similarly, bacterial biofilms, through their dynamic matrix properties, govern bacterial community formation, resistance to external stresses, and biofilm-associated infections~\cite{flemming2023biofilm}. Ignoring the dynamic nature of the environment could oversimplify the complexities of cell and bacterial behaviors, hindering our ability to comprehend processes like wound healing, cancer metastasis, and microbial infections. Hence, integrating the understanding of dynamic environmental changes is crucial for unraveling the intricate dynamics of active materials in biological systems.

\section{Author contributions}
J.I.-M and F.S. designed the experiments and A.D. designed the model.
I.V.-C. performed the experiments and extracted the director field and velocity field data.
A.A implemented the model, performed the simulations, analyzed the simulations and experimental data, and prepared figures with help from I.V.-C. and M.C.P.,
J.I.-M., F.S and A.D., all contributed to the interpretation of results.
A.D. prepared the first draft. A.D. and A.A. wrote the paper with input from all authors.
This collaborative effort was led by J.I.-M, F.S and A.D.

\section{Acknowledgements}
We thank Kristian Thijssen for the helpful discussions. A.A. acknowledges support from the EU’s Horizon Europe research and innovation program under the Marie Sklodowska-Curie grant agreement No. 101063870 (TopCellComm). A.D. acknowledges funding from the Novo Nordisk Foundation (grant no. NNF18SA0035142 and NERD grant no. NNF21OC0068687), Villum Fonden Grant no. 29476, and the European Union via the ERC-Starting Grant PhysCoMeT, grant no. 101041418. We thank M. Pons, A. LeRoux, and G. Iruela (Universitat de Barcelona) for their assistance in the expression of motor proteins.  I.V.-C. acknowledges funding from Generalitat de Catalunya through a FI-2020 PhD. Fellowship. I.V.-C., J.I.-M., and F.S. acknowledge funding from MICIU/AEI/10.13039/501100011033 (Grant No. PID2022-137713NB-C21). I.V.-C. and J.I.-M. acknowledge funding from MICIU/AEI/10.13039/501100011033 (Grant No. PDC2022-133625-I00). \ed{The authors are indebted to the Brandeis University MRSEC Biosynthesis facility, supported by NSF DMR-MRSEC 2011846, for providing the tubulin.}

\bibliography{apssamp}

\providecommand{\noopsort}[1]{}\providecommand{\singleletter}[1]{#1}%
\begin{thebibliography}{52}%
\makeatletter
\providecommand \@ifxundefined [1]{%
 \@ifx{#1\undefined}
}%
\providecommand \@ifnum [1]{%
 \ifnum #1\expandafter \@firstoftwo
 \else \expandafter \@secondoftwo
 \fi
}%
\providecommand \@ifx [1]{%
 \ifx #1\expandafter \@firstoftwo
 \else \expandafter \@secondoftwo
 \fi
}%
\providecommand \natexlab [1]{#1}%
\providecommand \enquote  [1]{``#1''}%
\providecommand \bibnamefont  [1]{#1}%
\providecommand \bibfnamefont [1]{#1}%
\providecommand \citenamefont [1]{#1}%
\providecommand \href@noop [0]{\@secondoftwo}%
\providecommand \href [0]{\begingroup \@sanitize@url \@href}%
\providecommand \@href[1]{\@@startlink{#1}\@@href}%
\providecommand \@@href[1]{\endgroup#1\@@endlink}%
\providecommand \@sanitize@url [0]{\catcode `\\12\catcode `\$12\catcode
  `\&12\catcode `\#12\catcode `\^12\catcode `\_12\catcode `\%12\relax}%
\providecommand \@@startlink[1]{}%
\providecommand \@@endlink[0]{}%
\providecommand \url  [0]{\begingroup\@sanitize@url \@url }%
\providecommand \@url [1]{\endgroup\@href {#1}{\urlprefix }}%
\providecommand \urlprefix  [0]{URL }%
\providecommand \Eprint [0]{\href }%
\providecommand \doibase [0]{https://doi.org/}%
\providecommand \selectlanguage [0]{\@gobble}%
\providecommand \bibinfo  [0]{\@secondoftwo}%
\providecommand \bibfield  [0]{\@secondoftwo}%
\providecommand \translation [1]{[#1]}%
\providecommand \BibitemOpen [0]{}%
\providecommand \bibitemStop [0]{}%
\providecommand \bibitemNoStop [0]{.\EOS\space}%
\providecommand \EOS [0]{\spacefactor3000\relax}%
\providecommand \BibitemShut  [1]{\csname bibitem#1\endcsname}%
\let\auto@bib@innerbib\@empty
\bibitem [{\citenamefont {Prost}\ \emph {et~al.}(2015)\citenamefont {Prost},
  \citenamefont {J{\"u}licher},\ and\ \citenamefont
  {Joanny}}]{prost2015active}%
  \BibitemOpen
  \bibfield  {author} {\bibinfo {author} {\bibfnamefont {J.}~\bibnamefont
  {Prost}}, \bibinfo {author} {\bibfnamefont {F.}~\bibnamefont
  {J{\"u}licher}},\ and\ \bibinfo {author} {\bibfnamefont {J.-F.}\ \bibnamefont
  {Joanny}},\ }\href@noop {} {\bibfield  {journal} {\bibinfo  {journal} {Nature
  Physics}\ }\textbf {\bibinfo {volume} {11}},\ \bibinfo {pages} {111}
  (\bibinfo {year} {2015})}\BibitemShut {NoStop}%
\bibitem [{\citenamefont {Hartmann}\ \emph {et~al.}(2019)\citenamefont
  {Hartmann}, \citenamefont {Singh}, \citenamefont {Pearce}, \citenamefont
  {Mok}, \citenamefont {Song}, \citenamefont {D{\'\i}az-Pascual}, \citenamefont
  {Dunkel},\ and\ \citenamefont {Drescher}}]{hartmann2019emergence}%
  \BibitemOpen
  \bibfield  {author} {\bibinfo {author} {\bibfnamefont {R.}~\bibnamefont
  {Hartmann}}, \bibinfo {author} {\bibfnamefont {P.~K.}\ \bibnamefont {Singh}},
  \bibinfo {author} {\bibfnamefont {P.}~\bibnamefont {Pearce}}, \bibinfo
  {author} {\bibfnamefont {R.}~\bibnamefont {Mok}}, \bibinfo {author}
  {\bibfnamefont {B.}~\bibnamefont {Song}}, \bibinfo {author} {\bibfnamefont
  {F.}~\bibnamefont {D{\'\i}az-Pascual}}, \bibinfo {author} {\bibfnamefont
  {J.}~\bibnamefont {Dunkel}},\ and\ \bibinfo {author} {\bibfnamefont
  {K.}~\bibnamefont {Drescher}},\ }\href@noop {} {\bibfield  {journal}
  {\bibinfo  {journal} {Nature Physics}\ }\textbf {\bibinfo {volume} {15}},\
  \bibinfo {pages} {251} (\bibinfo {year} {2019})}\BibitemShut {NoStop}%
\bibitem [{\citenamefont {Needleman}\ and\ \citenamefont
  {Dogic}(2017)}]{needleman2017active}%
  \BibitemOpen
  \bibfield  {author} {\bibinfo {author} {\bibfnamefont {D.}~\bibnamefont
  {Needleman}}\ and\ \bibinfo {author} {\bibfnamefont {Z.}~\bibnamefont
  {Dogic}},\ }\href@noop {} {\bibfield  {journal} {\bibinfo  {journal} {Nature
  Reviews Materials}\ }\textbf {\bibinfo {volume} {2}},\ \bibinfo {pages} {1}
  (\bibinfo {year} {2017})}\BibitemShut {NoStop}%
\bibitem [{\citenamefont {Doostmohammadi}\ \emph {et~al.}(2018)\citenamefont
  {Doostmohammadi}, \citenamefont {Ign{\'e}s-Mullol}, \citenamefont {Yeomans},\
  and\ \citenamefont {Sagu{\'e}s}}]{doostmohammadi2018active}%
  \BibitemOpen
  \bibfield  {author} {\bibinfo {author} {\bibfnamefont {A.}~\bibnamefont
  {Doostmohammadi}}, \bibinfo {author} {\bibfnamefont {J.}~\bibnamefont
  {Ign{\'e}s-Mullol}}, \bibinfo {author} {\bibfnamefont {J.~M.}\ \bibnamefont
  {Yeomans}},\ and\ \bibinfo {author} {\bibfnamefont {F.}~\bibnamefont
  {Sagu{\'e}s}},\ }\href@noop {} {\bibfield  {journal} {\bibinfo  {journal}
  {Nature Communications}\ }\textbf {\bibinfo {volume} {9}},\ \bibinfo {pages}
  {3246} (\bibinfo {year} {2018})}\BibitemShut {NoStop}%
\bibitem [{\citenamefont {Sagu{\'{e}}s}(2023)}]{sagues23}%
  \BibitemOpen
  \bibfield  {author} {\bibinfo {author} {\bibfnamefont {F.}~\bibnamefont
  {Sagu{\'{e}}s}},\ }\href@noop {} {\emph {\bibinfo {title} {Colloidal Active
  Matter}}}\ (\bibinfo  {publisher} {Taylor and Francis Group},\ \bibinfo
  {address} {Boca Raton, FL},\ \bibinfo {year} {2023})\BibitemShut {NoStop}%
\bibitem [{\citenamefont {Bianco}\ \emph {et~al.}(2007)\citenamefont {Bianco},
  \citenamefont {Poukkula}, \citenamefont {Cliffe}, \citenamefont {Mathieu},
  \citenamefont {Luque}, \citenamefont {Fulga},\ and\ \citenamefont
  {R{\o}rth}}]{bianco2007two}%
  \BibitemOpen
  \bibfield  {author} {\bibinfo {author} {\bibfnamefont {A.}~\bibnamefont
  {Bianco}}, \bibinfo {author} {\bibfnamefont {M.}~\bibnamefont {Poukkula}},
  \bibinfo {author} {\bibfnamefont {A.}~\bibnamefont {Cliffe}}, \bibinfo
  {author} {\bibfnamefont {J.}~\bibnamefont {Mathieu}}, \bibinfo {author}
  {\bibfnamefont {C.~M.}\ \bibnamefont {Luque}}, \bibinfo {author}
  {\bibfnamefont {T.~A.}\ \bibnamefont {Fulga}},\ and\ \bibinfo {author}
  {\bibfnamefont {P.}~\bibnamefont {R{\o}rth}},\ }\href@noop {} {\bibfield
  {journal} {\bibinfo  {journal} {Nature}\ }\textbf {\bibinfo {volume} {448}},\
  \bibinfo {pages} {362} (\bibinfo {year} {2007})}\BibitemShut {NoStop}%
\bibitem [{\citenamefont {Maroudas-Sacks}\ \emph {et~al.}(2021)\citenamefont
  {Maroudas-Sacks}, \citenamefont {Garion}, \citenamefont {Shani-Zerbib},
  \citenamefont {Livshits}, \citenamefont {Braun},\ and\ \citenamefont
  {Keren}}]{maroudas2021topological}%
  \BibitemOpen
  \bibfield  {author} {\bibinfo {author} {\bibfnamefont {Y.}~\bibnamefont
  {Maroudas-Sacks}}, \bibinfo {author} {\bibfnamefont {L.}~\bibnamefont
  {Garion}}, \bibinfo {author} {\bibfnamefont {L.}~\bibnamefont
  {Shani-Zerbib}}, \bibinfo {author} {\bibfnamefont {A.}~\bibnamefont
  {Livshits}}, \bibinfo {author} {\bibfnamefont {E.}~\bibnamefont {Braun}},\
  and\ \bibinfo {author} {\bibfnamefont {K.}~\bibnamefont {Keren}},\
  }\href@noop {} {\bibfield  {journal} {\bibinfo  {journal} {Nature Physics}\
  }\textbf {\bibinfo {volume} {17}},\ \bibinfo {pages} {251} (\bibinfo {year}
  {2021})}\BibitemShut {NoStop}%
\bibitem [{\citenamefont {Ariel}\ \emph {et~al.}(2015)\citenamefont {Ariel},
  \citenamefont {Rabani}, \citenamefont {Benisty}, \citenamefont {Partridge},
  \citenamefont {Harshey},\ and\ \citenamefont {Be'Er}}]{ariel2015swarming}%
  \BibitemOpen
  \bibfield  {author} {\bibinfo {author} {\bibfnamefont {G.}~\bibnamefont
  {Ariel}}, \bibinfo {author} {\bibfnamefont {A.}~\bibnamefont {Rabani}},
  \bibinfo {author} {\bibfnamefont {S.}~\bibnamefont {Benisty}}, \bibinfo
  {author} {\bibfnamefont {J.~D.}\ \bibnamefont {Partridge}}, \bibinfo {author}
  {\bibfnamefont {R.~M.}\ \bibnamefont {Harshey}},\ and\ \bibinfo {author}
  {\bibfnamefont {A.}~\bibnamefont {Be'Er}},\ }\href@noop {} {\bibfield
  {journal} {\bibinfo  {journal} {Nature Communications}\ }\textbf {\bibinfo
  {volume} {6}},\ \bibinfo {pages} {8396} (\bibinfo {year} {2015})}\BibitemShut
  {NoStop}%
\bibitem [{\citenamefont {Clark}\ and\ \citenamefont
  {Vignjevic}(2015)}]{clark2015modes}%
  \BibitemOpen
  \bibfield  {author} {\bibinfo {author} {\bibfnamefont {A.~G.}\ \bibnamefont
  {Clark}}\ and\ \bibinfo {author} {\bibfnamefont {D.~M.}\ \bibnamefont
  {Vignjevic}},\ }\href@noop {} {\bibfield  {journal} {\bibinfo  {journal}
  {Current Opinion in Cell Biology}\ }\textbf {\bibinfo {volume} {36}},\
  \bibinfo {pages} {13} (\bibinfo {year} {2015})}\BibitemShut {NoStop}%
\bibitem [{\citenamefont {Zhang}\ \emph {et~al.}(2022)\citenamefont {Zhang},
  \citenamefont {Chen}, \citenamefont {Li}, \citenamefont {Zhang},\ and\
  \citenamefont {Li}}]{zhang2022topological}%
  \BibitemOpen
  \bibfield  {author} {\bibinfo {author} {\bibfnamefont {D.-Q.}\ \bibnamefont
  {Zhang}}, \bibinfo {author} {\bibfnamefont {P.-C.}\ \bibnamefont {Chen}},
  \bibinfo {author} {\bibfnamefont {Z.-Y.}\ \bibnamefont {Li}}, \bibinfo
  {author} {\bibfnamefont {R.}~\bibnamefont {Zhang}},\ and\ \bibinfo {author}
  {\bibfnamefont {B.}~\bibnamefont {Li}},\ }\href@noop {} {\bibfield  {journal}
  {\bibinfo  {journal} {Proceedings of the National Academy of Sciences}\
  }\textbf {\bibinfo {volume} {119}},\ \bibinfo {pages} {e2122494119} (\bibinfo
  {year} {2022})}\BibitemShut {NoStop}%
\bibitem [{\citenamefont {Sanchez}\ \emph {et~al.}(2012)\citenamefont
  {Sanchez}, \citenamefont {Chen}, \citenamefont {DeCamp}, \citenamefont
  {Heymann},\ and\ \citenamefont {Dogic}}]{sanchez2012spontaneous}%
  \BibitemOpen
  \bibfield  {author} {\bibinfo {author} {\bibfnamefont {T.}~\bibnamefont
  {Sanchez}}, \bibinfo {author} {\bibfnamefont {D.~T.}\ \bibnamefont {Chen}},
  \bibinfo {author} {\bibfnamefont {S.~J.}\ \bibnamefont {DeCamp}}, \bibinfo
  {author} {\bibfnamefont {M.}~\bibnamefont {Heymann}},\ and\ \bibinfo {author}
  {\bibfnamefont {Z.}~\bibnamefont {Dogic}},\ }\href@noop {} {\bibfield
  {journal} {\bibinfo  {journal} {Nature}\ }\textbf {\bibinfo {volume} {491}},\
  \bibinfo {pages} {431} (\bibinfo {year} {2012})}\BibitemShut {NoStop}%
\bibitem [{\citenamefont {Schaller}\ \emph {et~al.}(2010)\citenamefont
  {Schaller}, \citenamefont {Weber}, \citenamefont {Semmrich}, \citenamefont
  {Frey},\ and\ \citenamefont {Bausch}}]{schaller2010polar}%
  \BibitemOpen
  \bibfield  {author} {\bibinfo {author} {\bibfnamefont {V.}~\bibnamefont
  {Schaller}}, \bibinfo {author} {\bibfnamefont {C.}~\bibnamefont {Weber}},
  \bibinfo {author} {\bibfnamefont {C.}~\bibnamefont {Semmrich}}, \bibinfo
  {author} {\bibfnamefont {E.}~\bibnamefont {Frey}},\ and\ \bibinfo {author}
  {\bibfnamefont {A.~R.}\ \bibnamefont {Bausch}},\ }\href@noop {} {\bibfield
  {journal} {\bibinfo  {journal} {Nature}\ }\textbf {\bibinfo {volume} {467}},\
  \bibinfo {pages} {73} (\bibinfo {year} {2010})}\BibitemShut {NoStop}%
\bibitem [{\citenamefont {Ramaswamy}(2010)}]{ramaswamy2010mechanics}%
  \BibitemOpen
  \bibfield  {author} {\bibinfo {author} {\bibfnamefont {S.}~\bibnamefont
  {Ramaswamy}},\ }\href@noop {} {\bibfield  {journal} {\bibinfo  {journal}
  {Annual Review of Condensed Matter Physics}\ }\textbf {\bibinfo {volume}
  {1}},\ \bibinfo {pages} {323} (\bibinfo {year} {2010})}\BibitemShut {NoStop}%
\bibitem [{\citenamefont {Marchetti}\ \emph {et~al.}(2013)\citenamefont
  {Marchetti}, \citenamefont {Joanny}, \citenamefont {Ramaswamy}, \citenamefont
  {Liverpool}, \citenamefont {Prost}, \citenamefont {Rao},\ and\ \citenamefont
  {Simha}}]{marchetti2013hydrodynamics}%
  \BibitemOpen
  \bibfield  {author} {\bibinfo {author} {\bibfnamefont {M.~C.}\ \bibnamefont
  {Marchetti}}, \bibinfo {author} {\bibfnamefont {J.-F.}\ \bibnamefont
  {Joanny}}, \bibinfo {author} {\bibfnamefont {S.}~\bibnamefont {Ramaswamy}},
  \bibinfo {author} {\bibfnamefont {T.~B.}\ \bibnamefont {Liverpool}}, \bibinfo
  {author} {\bibfnamefont {J.}~\bibnamefont {Prost}}, \bibinfo {author}
  {\bibfnamefont {M.}~\bibnamefont {Rao}},\ and\ \bibinfo {author}
  {\bibfnamefont {R.~A.}\ \bibnamefont {Simha}},\ }\href@noop {} {\bibfield
  {journal} {\bibinfo  {journal} {Reviews of Modern Physics}\ }\textbf
  {\bibinfo {volume} {85}},\ \bibinfo {pages} {1143} (\bibinfo {year}
  {2013})}\BibitemShut {NoStop}%
\bibitem [{\citenamefont {Alert}\ \emph {et~al.}(2022)\citenamefont {Alert},
  \citenamefont {Casademunt},\ and\ \citenamefont {Joanny}}]{alert2022active}%
  \BibitemOpen
  \bibfield  {author} {\bibinfo {author} {\bibfnamefont {R.}~\bibnamefont
  {Alert}}, \bibinfo {author} {\bibfnamefont {J.}~\bibnamefont {Casademunt}},\
  and\ \bibinfo {author} {\bibfnamefont {J.-F.}\ \bibnamefont {Joanny}},\
  }\href@noop {} {\bibfield  {journal} {\bibinfo  {journal} {Annual Review of
  Condensed Matter Physics}\ }\textbf {\bibinfo {volume} {13}},\ \bibinfo
  {pages} {143} (\bibinfo {year} {2022})}\BibitemShut {NoStop}%
\bibitem [{\citenamefont {Doostmohammadi}\ and\ \citenamefont
  {Ladoux}(2022)}]{doostmohammadi2022physics}%
  \BibitemOpen
  \bibfield  {author} {\bibinfo {author} {\bibfnamefont {A.}~\bibnamefont
  {Doostmohammadi}}\ and\ \bibinfo {author} {\bibfnamefont {B.}~\bibnamefont
  {Ladoux}},\ }\href@noop {} {\bibfield  {journal} {\bibinfo  {journal} {Trends
  in Cell Biology}\ }\textbf {\bibinfo {volume} {32}},\ \bibinfo {pages} {140}
  (\bibinfo {year} {2022})}\BibitemShut {NoStop}%
\bibitem [{\citenamefont {Callan-Jones}\ and\ \citenamefont
  {Voituriez}(2016)}]{callan2016actin}%
  \BibitemOpen
  \bibfield  {author} {\bibinfo {author} {\bibfnamefont {A.~C.}\ \bibnamefont
  {Callan-Jones}}\ and\ \bibinfo {author} {\bibfnamefont {R.}~\bibnamefont
  {Voituriez}},\ }\href@noop {} {\bibfield  {journal} {\bibinfo  {journal}
  {Current Opinion in Cell Biology}\ }\textbf {\bibinfo {volume} {38}},\
  \bibinfo {pages} {12} (\bibinfo {year} {2016})}\BibitemShut {NoStop}%
\bibitem [{\citenamefont {Gomez}\ \emph {et~al.}(2023)\citenamefont {Gomez},
  \citenamefont {Bureau}, \citenamefont {John}, \citenamefont {Ch{\^e}ne},
  \citenamefont {D{\'e}barre},\ and\ \citenamefont
  {Lecuyer}}]{gomez2023substrate}%
  \BibitemOpen
  \bibfield  {author} {\bibinfo {author} {\bibfnamefont {S.}~\bibnamefont
  {Gomez}}, \bibinfo {author} {\bibfnamefont {L.}~\bibnamefont {Bureau}},
  \bibinfo {author} {\bibfnamefont {K.}~\bibnamefont {John}}, \bibinfo {author}
  {\bibfnamefont {E.-N.}\ \bibnamefont {Ch{\^e}ne}}, \bibinfo {author}
  {\bibfnamefont {D.}~\bibnamefont {D{\'e}barre}},\ and\ \bibinfo {author}
  {\bibfnamefont {S.}~\bibnamefont {Lecuyer}},\ }\href@noop {} {\bibfield
  {journal} {\bibinfo  {journal} {Elife}\ }\textbf {\bibinfo {volume} {12}},\
  \bibinfo {pages} {e81112} (\bibinfo {year} {2023})}\BibitemShut {NoStop}%
\bibitem [{\citenamefont {Krishnan}\ \emph {et~al.}(2011)\citenamefont
  {Krishnan}, \citenamefont {Klumpers}, \citenamefont {Park}, \citenamefont
  {Rajendran}, \citenamefont {Trepat}, \citenamefont {Van~Bezu}, \citenamefont
  {Van~Hinsbergh}, \citenamefont {Carman}, \citenamefont {Brain}, \citenamefont
  {Fredberg} \emph {et~al.}}]{krishnan2011substrate}%
  \BibitemOpen
  \bibfield  {author} {\bibinfo {author} {\bibfnamefont {R.}~\bibnamefont
  {Krishnan}}, \bibinfo {author} {\bibfnamefont {D.~D.}\ \bibnamefont
  {Klumpers}}, \bibinfo {author} {\bibfnamefont {C.~Y.}\ \bibnamefont {Park}},
  \bibinfo {author} {\bibfnamefont {K.}~\bibnamefont {Rajendran}}, \bibinfo
  {author} {\bibfnamefont {X.}~\bibnamefont {Trepat}}, \bibinfo {author}
  {\bibfnamefont {J.}~\bibnamefont {Van~Bezu}}, \bibinfo {author}
  {\bibfnamefont {V.~W.}\ \bibnamefont {Van~Hinsbergh}}, \bibinfo {author}
  {\bibfnamefont {C.~V.}\ \bibnamefont {Carman}}, \bibinfo {author}
  {\bibfnamefont {J.~D.}\ \bibnamefont {Brain}}, \bibinfo {author}
  {\bibfnamefont {J.~J.}\ \bibnamefont {Fredberg}}, \emph {et~al.},\
  }\href@noop {} {\bibfield  {journal} {\bibinfo  {journal} {American Journal
  of Physiology-Cell Physiology}\ }\textbf {\bibinfo {volume} {300}},\ \bibinfo
  {pages} {C146} (\bibinfo {year} {2011})}\BibitemShut {NoStop}%
\bibitem [{\citenamefont {Doostmohammadi}\ \emph {et~al.}(2017)\citenamefont
  {Doostmohammadi}, \citenamefont {Shendruk}, \citenamefont {Thijssen},\ and\
  \citenamefont {Yeomans}}]{doostmohammadi2017onset}%
  \BibitemOpen
  \bibfield  {author} {\bibinfo {author} {\bibfnamefont {A.}~\bibnamefont
  {Doostmohammadi}}, \bibinfo {author} {\bibfnamefont {T.~N.}\ \bibnamefont
  {Shendruk}}, \bibinfo {author} {\bibfnamefont {K.}~\bibnamefont {Thijssen}},\
  and\ \bibinfo {author} {\bibfnamefont {J.~M.}\ \bibnamefont {Yeomans}},\
  }\href@noop {} {\bibfield  {journal} {\bibinfo  {journal} {Nature
  Communications}\ }\textbf {\bibinfo {volume} {8}},\ \bibinfo {pages} {15326}
  (\bibinfo {year} {2017})}\BibitemShut {NoStop}%
\bibitem [{\citenamefont {Shendruk}\ \emph {et~al.}(2017)\citenamefont
  {Shendruk}, \citenamefont {Doostmohammadi}, \citenamefont {Thijssen},\ and\
  \citenamefont {Yeomans}}]{shendruk2017dancing}%
  \BibitemOpen
  \bibfield  {author} {\bibinfo {author} {\bibfnamefont {T.~N.}\ \bibnamefont
  {Shendruk}}, \bibinfo {author} {\bibfnamefont {A.}~\bibnamefont
  {Doostmohammadi}}, \bibinfo {author} {\bibfnamefont {K.}~\bibnamefont
  {Thijssen}},\ and\ \bibinfo {author} {\bibfnamefont {J.~M.}\ \bibnamefont
  {Yeomans}},\ }\href@noop {} {\bibfield  {journal} {\bibinfo  {journal} {Soft
  Matter}\ }\textbf {\bibinfo {volume} {13}},\ \bibinfo {pages} {3853}
  (\bibinfo {year} {2017})}\BibitemShut {NoStop}%
\bibitem [{\citenamefont {Samui}\ \emph {et~al.}(2021)\citenamefont {Samui},
  \citenamefont {Yeomans},\ and\ \citenamefont {Thampi}}]{samui2021flow}%
  \BibitemOpen
  \bibfield  {author} {\bibinfo {author} {\bibfnamefont {A.}~\bibnamefont
  {Samui}}, \bibinfo {author} {\bibfnamefont {J.~M.}\ \bibnamefont {Yeomans}},\
  and\ \bibinfo {author} {\bibfnamefont {S.~P.}\ \bibnamefont {Thampi}},\
  }\href@noop {} {\bibfield  {journal} {\bibinfo  {journal} {Soft Matter}\
  }\textbf {\bibinfo {volume} {17}},\ \bibinfo {pages} {10640} (\bibinfo {year}
  {2021})}\BibitemShut {NoStop}%
\bibitem [{\citenamefont {Coelho}\ \emph {et~al.}(2019)\citenamefont {Coelho},
  \citenamefont {Ara{\'u}jo},\ and\ \citenamefont
  {da~Gama}}]{coelho2019active}%
  \BibitemOpen
  \bibfield  {author} {\bibinfo {author} {\bibfnamefont {R.~C.}\ \bibnamefont
  {Coelho}}, \bibinfo {author} {\bibfnamefont {N.~A.}\ \bibnamefont
  {Ara{\'u}jo}},\ and\ \bibinfo {author} {\bibfnamefont {M.~M.~T.}\
  \bibnamefont {da~Gama}},\ }\href@noop {} {\bibfield  {journal} {\bibinfo
  {journal} {Soft Matter}\ }\textbf {\bibinfo {volume} {15}},\ \bibinfo {pages}
  {6819} (\bibinfo {year} {2019})}\BibitemShut {NoStop}%
\bibitem [{\citenamefont {Hardo{\"u}in}\ \emph {et~al.}(2019)\citenamefont
  {Hardo{\"u}in}, \citenamefont {Hughes}, \citenamefont {Doostmohammadi},
  \citenamefont {Laurent}, \citenamefont {Lopez-Leon}, \citenamefont {Yeomans},
  \citenamefont {Ign{\'e}s-Mullol},\ and\ \citenamefont
  {Sagu{\'e}s}}]{hardouin2019reconfigurable}%
  \BibitemOpen
  \bibfield  {author} {\bibinfo {author} {\bibfnamefont {J.}~\bibnamefont
  {Hardo{\"u}in}}, \bibinfo {author} {\bibfnamefont {R.}~\bibnamefont
  {Hughes}}, \bibinfo {author} {\bibfnamefont {A.}~\bibnamefont
  {Doostmohammadi}}, \bibinfo {author} {\bibfnamefont {J.}~\bibnamefont
  {Laurent}}, \bibinfo {author} {\bibfnamefont {T.}~\bibnamefont {Lopez-Leon}},
  \bibinfo {author} {\bibfnamefont {J.~M.}\ \bibnamefont {Yeomans}}, \bibinfo
  {author} {\bibfnamefont {J.}~\bibnamefont {Ign{\'e}s-Mullol}},\ and\ \bibinfo
  {author} {\bibfnamefont {F.}~\bibnamefont {Sagu{\'e}s}},\ }\href@noop {}
  {\bibfield  {journal} {\bibinfo  {journal} {Communications Physics}\ }\textbf
  {\bibinfo {volume} {2}},\ \bibinfo {pages} {121} (\bibinfo {year}
  {2019})}\BibitemShut {NoStop}%
\bibitem [{\citenamefont {Opathalage}\ \emph {et~al.}(2019)\citenamefont
  {Opathalage}, \citenamefont {Norton}, \citenamefont {Juniper}, \citenamefont
  {Langeslay}, \citenamefont {Aghvami}, \citenamefont {Fraden},\ and\
  \citenamefont {Dogic}}]{opathalage2019self}%
  \BibitemOpen
  \bibfield  {author} {\bibinfo {author} {\bibfnamefont {A.}~\bibnamefont
  {Opathalage}}, \bibinfo {author} {\bibfnamefont {M.~M.}\ \bibnamefont
  {Norton}}, \bibinfo {author} {\bibfnamefont {M.~P.}\ \bibnamefont {Juniper}},
  \bibinfo {author} {\bibfnamefont {B.}~\bibnamefont {Langeslay}}, \bibinfo
  {author} {\bibfnamefont {S.~A.}\ \bibnamefont {Aghvami}}, \bibinfo {author}
  {\bibfnamefont {S.}~\bibnamefont {Fraden}},\ and\ \bibinfo {author}
  {\bibfnamefont {Z.}~\bibnamefont {Dogic}},\ }\href@noop {} {\bibfield
  {journal} {\bibinfo  {journal} {Proceedings of the National Academy of
  Sciences}\ }\textbf {\bibinfo {volume} {116}},\ \bibinfo {pages} {4788}
  (\bibinfo {year} {2019})}\BibitemShut {NoStop}%
\bibitem [{\citenamefont {Hardo{\"u}in}\ \emph {et~al.}(2022)\citenamefont
  {Hardo{\"u}in}, \citenamefont {Dor{\'e}}, \citenamefont {Laurent},
  \citenamefont {Lopez-Leon}, \citenamefont {Ign{\'e}s-Mullol},\ and\
  \citenamefont {Sagu{\'e}s}}]{hardouin2022active}%
  \BibitemOpen
  \bibfield  {author} {\bibinfo {author} {\bibfnamefont {J.}~\bibnamefont
  {Hardo{\"u}in}}, \bibinfo {author} {\bibfnamefont {C.}~\bibnamefont
  {Dor{\'e}}}, \bibinfo {author} {\bibfnamefont {J.}~\bibnamefont {Laurent}},
  \bibinfo {author} {\bibfnamefont {T.}~\bibnamefont {Lopez-Leon}}, \bibinfo
  {author} {\bibfnamefont {J.}~\bibnamefont {Ign{\'e}s-Mullol}},\ and\ \bibinfo
  {author} {\bibfnamefont {F.}~\bibnamefont {Sagu{\'e}s}},\ }\href@noop {}
  {\bibfield  {journal} {\bibinfo  {journal} {Nature Communications}\ }\textbf
  {\bibinfo {volume} {13}},\ \bibinfo {pages} {6675} (\bibinfo {year}
  {2022})}\BibitemShut {NoStop}%
\bibitem [{\citenamefont {Lushi}\ \emph {et~al.}(2014)\citenamefont {Lushi},
  \citenamefont {Wioland},\ and\ \citenamefont {Goldstein}}]{lushi2014fluid}%
  \BibitemOpen
  \bibfield  {author} {\bibinfo {author} {\bibfnamefont {E.}~\bibnamefont
  {Lushi}}, \bibinfo {author} {\bibfnamefont {H.}~\bibnamefont {Wioland}},\
  and\ \bibinfo {author} {\bibfnamefont {R.~E.}\ \bibnamefont {Goldstein}},\
  }\href@noop {} {\bibfield  {journal} {\bibinfo  {journal} {Proceedings of the
  National Academy of Sciences}\ }\textbf {\bibinfo {volume} {111}},\ \bibinfo
  {pages} {9733} (\bibinfo {year} {2014})}\BibitemShut {NoStop}%
\bibitem [{\citenamefont {You}\ \emph {et~al.}(2021)\citenamefont {You},
  \citenamefont {Pearce},\ and\ \citenamefont {Giomi}}]{you2021confinement}%
  \BibitemOpen
  \bibfield  {author} {\bibinfo {author} {\bibfnamefont {Z.}~\bibnamefont
  {You}}, \bibinfo {author} {\bibfnamefont {D.~J.}\ \bibnamefont {Pearce}},\
  and\ \bibinfo {author} {\bibfnamefont {L.}~\bibnamefont {Giomi}},\
  }\href@noop {} {\bibfield  {journal} {\bibinfo  {journal} {Science Advances}\
  }\textbf {\bibinfo {volume} {7}},\ \bibinfo {pages} {eabc8685} (\bibinfo
  {year} {2021})}\BibitemShut {NoStop}%
\bibitem [{\citenamefont {Deforet}\ \emph {et~al.}(2014)\citenamefont
  {Deforet}, \citenamefont {Hakim}, \citenamefont {Yevick}, \citenamefont
  {Duclos},\ and\ \citenamefont {Silberzan}}]{deforet2014emergence}%
  \BibitemOpen
  \bibfield  {author} {\bibinfo {author} {\bibfnamefont {M.}~\bibnamefont
  {Deforet}}, \bibinfo {author} {\bibfnamefont {V.}~\bibnamefont {Hakim}},
  \bibinfo {author} {\bibfnamefont {H.~G.}\ \bibnamefont {Yevick}}, \bibinfo
  {author} {\bibfnamefont {G.}~\bibnamefont {Duclos}},\ and\ \bibinfo {author}
  {\bibfnamefont {P.}~\bibnamefont {Silberzan}},\ }\href@noop {} {\bibfield
  {journal} {\bibinfo  {journal} {Nature Communications}\ }\textbf {\bibinfo
  {volume} {5}},\ \bibinfo {pages} {3747} (\bibinfo {year} {2014})}\BibitemShut
  {NoStop}%
\bibitem [{\citenamefont {Duclos}\ \emph {et~al.}(2018)\citenamefont {Duclos},
  \citenamefont {Blanch-Mercader}, \citenamefont {Yashunsky}, \citenamefont
  {Salbreux}, \citenamefont {Joanny}, \citenamefont {Prost},\ and\
  \citenamefont {Silberzan}}]{duclos2018spontaneous}%
  \BibitemOpen
  \bibfield  {author} {\bibinfo {author} {\bibfnamefont {G.}~\bibnamefont
  {Duclos}}, \bibinfo {author} {\bibfnamefont {C.}~\bibnamefont
  {Blanch-Mercader}}, \bibinfo {author} {\bibfnamefont {V.}~\bibnamefont
  {Yashunsky}}, \bibinfo {author} {\bibfnamefont {G.}~\bibnamefont {Salbreux}},
  \bibinfo {author} {\bibfnamefont {J.-F.}\ \bibnamefont {Joanny}}, \bibinfo
  {author} {\bibfnamefont {J.}~\bibnamefont {Prost}},\ and\ \bibinfo {author}
  {\bibfnamefont {P.}~\bibnamefont {Silberzan}},\ }\href@noop {} {\bibfield
  {journal} {\bibinfo  {journal} {Nature Physics}\ }\textbf {\bibinfo {volume}
  {14}},\ \bibinfo {pages} {728} (\bibinfo {year} {2018})}\BibitemShut
  {NoStop}%
\bibitem [{\citenamefont {Thampi}\ \emph {et~al.}(2014)\citenamefont {Thampi},
  \citenamefont {Golestanian},\ and\ \citenamefont
  {Yeomans}}]{thampi2014active}%
  \BibitemOpen
  \bibfield  {author} {\bibinfo {author} {\bibfnamefont {S.~P.}\ \bibnamefont
  {Thampi}}, \bibinfo {author} {\bibfnamefont {R.}~\bibnamefont
  {Golestanian}},\ and\ \bibinfo {author} {\bibfnamefont {J.~M.}\ \bibnamefont
  {Yeomans}},\ }\href@noop {} {\bibfield  {journal} {\bibinfo  {journal}
  {Physical Review E}\ }\textbf {\bibinfo {volume} {90}},\ \bibinfo {pages}
  {062307} (\bibinfo {year} {2014})}\BibitemShut {NoStop}%
\bibitem [{\citenamefont {Thijssen}\ \emph {et~al.}(2020)\citenamefont
  {Thijssen}, \citenamefont {Nejad},\ and\ \citenamefont
  {Yeomans}}]{thijssen2020role}%
  \BibitemOpen
  \bibfield  {author} {\bibinfo {author} {\bibfnamefont {K.}~\bibnamefont
  {Thijssen}}, \bibinfo {author} {\bibfnamefont {M.~R.}\ \bibnamefont
  {Nejad}},\ and\ \bibinfo {author} {\bibfnamefont {J.~M.}\ \bibnamefont
  {Yeomans}},\ }\href@noop {} {\bibfield  {journal} {\bibinfo  {journal}
  {Physical Review Letters}\ }\textbf {\bibinfo {volume} {125}},\ \bibinfo
  {pages} {218004} (\bibinfo {year} {2020})}\BibitemShut {NoStop}%
\bibitem [{\citenamefont {Ramaswamy}\ \emph {et~al.}(2003)\citenamefont
  {Ramaswamy}, \citenamefont {Simha},\ and\ \citenamefont
  {Toner}}]{ramaswamy2003active}%
  \BibitemOpen
  \bibfield  {author} {\bibinfo {author} {\bibfnamefont {S.}~\bibnamefont
  {Ramaswamy}}, \bibinfo {author} {\bibfnamefont {R.~A.}\ \bibnamefont
  {Simha}},\ and\ \bibinfo {author} {\bibfnamefont {J.}~\bibnamefont {Toner}},\
  }\href@noop {} {\bibfield  {journal} {\bibinfo  {journal} {Europhysics
  Letters}\ }\textbf {\bibinfo {volume} {62}},\ \bibinfo {pages} {196}
  (\bibinfo {year} {2003})}\BibitemShut {NoStop}%
\bibitem [{\citenamefont {Thijssen}\ \emph {et~al.}(2021)\citenamefont
  {Thijssen}, \citenamefont {Khaladj}, \citenamefont {Aghvami}, \citenamefont
  {Gharbi}, \citenamefont {Fraden}, \citenamefont {Yeomans}, \citenamefont
  {Hirst},\ and\ \citenamefont {Shendruk}}]{thijssen2021submersed}%
  \BibitemOpen
  \bibfield  {author} {\bibinfo {author} {\bibfnamefont {K.}~\bibnamefont
  {Thijssen}}, \bibinfo {author} {\bibfnamefont {D.~A.}\ \bibnamefont
  {Khaladj}}, \bibinfo {author} {\bibfnamefont {S.~A.}\ \bibnamefont
  {Aghvami}}, \bibinfo {author} {\bibfnamefont {M.~A.}\ \bibnamefont {Gharbi}},
  \bibinfo {author} {\bibfnamefont {S.}~\bibnamefont {Fraden}}, \bibinfo
  {author} {\bibfnamefont {J.~M.}\ \bibnamefont {Yeomans}}, \bibinfo {author}
  {\bibfnamefont {L.~S.}\ \bibnamefont {Hirst}},\ and\ \bibinfo {author}
  {\bibfnamefont {T.~N.}\ \bibnamefont {Shendruk}},\ }\href@noop {} {\bibfield
  {journal} {\bibinfo  {journal} {Proceedings of the National Academy of
  Sciences}\ }\textbf {\bibinfo {volume} {118}},\ \bibinfo {pages}
  {e2106038118} (\bibinfo {year} {2021})}\BibitemShut {NoStop}%
\bibitem [{\citenamefont {Zhang}\ \emph {et~al.}(2021)\citenamefont {Zhang},
  \citenamefont {Redford}, \citenamefont {Ruijgrok}, \citenamefont {Kumar},
  \citenamefont {Mozaffari}, \citenamefont {Zemsky}, \citenamefont {Dinner},
  \citenamefont {Vitelli}, \citenamefont {Bryant}, \citenamefont {Gardel} \emph
  {et~al.}}]{zhang2021spatiotemporal}%
  \BibitemOpen
  \bibfield  {author} {\bibinfo {author} {\bibfnamefont {R.}~\bibnamefont
  {Zhang}}, \bibinfo {author} {\bibfnamefont {S.~A.}\ \bibnamefont {Redford}},
  \bibinfo {author} {\bibfnamefont {P.~V.}\ \bibnamefont {Ruijgrok}}, \bibinfo
  {author} {\bibfnamefont {N.}~\bibnamefont {Kumar}}, \bibinfo {author}
  {\bibfnamefont {A.}~\bibnamefont {Mozaffari}}, \bibinfo {author}
  {\bibfnamefont {S.}~\bibnamefont {Zemsky}}, \bibinfo {author} {\bibfnamefont
  {A.~R.}\ \bibnamefont {Dinner}}, \bibinfo {author} {\bibfnamefont
  {V.}~\bibnamefont {Vitelli}}, \bibinfo {author} {\bibfnamefont
  {Z.}~\bibnamefont {Bryant}}, \bibinfo {author} {\bibfnamefont {M.~L.}\
  \bibnamefont {Gardel}}, \emph {et~al.},\ }\href@noop {} {\bibfield  {journal}
  {\bibinfo  {journal} {Nature Materials}\ }\textbf {\bibinfo {volume} {20}},\
  \bibinfo {pages} {875} (\bibinfo {year} {2021})}\BibitemShut {NoStop}%
\bibitem [{\citenamefont {V{\'e}lez-Cer{\'o}n}\ \emph
  {et~al.}(2024)\citenamefont {V{\'e}lez-Cer{\'o}n}, \citenamefont {Guillamat},
  \citenamefont {Sagu{\'e}s},\ and\ \citenamefont
  {Ign{\'e}s-Mullol}}]{Velez24}%
  \BibitemOpen
  \bibfield  {author} {\bibinfo {author} {\bibfnamefont {I.}~\bibnamefont
  {V{\'e}lez-Cer{\'o}n}}, \bibinfo {author} {\bibfnamefont {P.}~\bibnamefont
  {Guillamat}}, \bibinfo {author} {\bibfnamefont {F.}~\bibnamefont
  {Sagu{\'e}s}},\ and\ \bibinfo {author} {\bibfnamefont {J.}~\bibnamefont
  {Ign{\'e}s-Mullol}},\ }\href@noop {} {\bibfield  {journal} {\bibinfo
  {journal} {Proceedings of the National Academy of Sciences}\ }\textbf
  {\bibinfo {volume} {121}},\ \bibinfo {pages} {e2312494121} (\bibinfo {year}
  {2024})}\BibitemShut {NoStop}%
\bibitem [{\citenamefont {Guillamat}\ \emph {et~al.}(2016)\citenamefont
  {Guillamat}, \citenamefont {Ign{\'e}s-Mullol},\ and\ \citenamefont
  {Sagu{\'e}s}}]{guillamat2016control}%
  \BibitemOpen
  \bibfield  {author} {\bibinfo {author} {\bibfnamefont {P.}~\bibnamefont
  {Guillamat}}, \bibinfo {author} {\bibfnamefont {J.}~\bibnamefont
  {Ign{\'e}s-Mullol}},\ and\ \bibinfo {author} {\bibfnamefont {F.}~\bibnamefont
  {Sagu{\'e}s}},\ }\href@noop {} {\bibfield  {journal} {\bibinfo  {journal}
  {Proceedings of the National Academy of Sciences}\ }\textbf {\bibinfo
  {volume} {113}},\ \bibinfo {pages} {5498} (\bibinfo {year}
  {2016})}\BibitemShut {NoStop}%
\bibitem [{sup()}]{supplementary}%
  \BibitemOpen
  \href@noop {} {\bibinfo {title} {See supplemental material at [link], which
  includes refs. \cite{blow2014biphasic, sultan2022quadrupolar,
  maitra2018nonequilibrium, cabral1993imaging, vanderWalt2014,
  copenhagen2021topological, thampi2014active,
  kawaguchi2017topological}}}\BibitemShut {NoStop}%
\bibitem [{\citenamefont {Blow}\ \emph {et~al.}(2014)\citenamefont {Blow},
  \citenamefont {Thampi},\ and\ \citenamefont {Yeomans}}]{blow2014biphasic}%
  \BibitemOpen
  \bibfield  {author} {\bibinfo {author} {\bibfnamefont {M.~L.}\ \bibnamefont
  {Blow}}, \bibinfo {author} {\bibfnamefont {S.~P.}\ \bibnamefont {Thampi}},\
  and\ \bibinfo {author} {\bibfnamefont {J.~M.}\ \bibnamefont {Yeomans}},\
  }\href@noop {} {\bibfield  {journal} {\bibinfo  {journal} {Physical Review
  Letters}\ }\textbf {\bibinfo {volume} {113}},\ \bibinfo {pages} {248303}
  (\bibinfo {year} {2014})}\BibitemShut {NoStop}%
\bibitem [{\citenamefont {Marenduzzo}\ \emph {et~al.}(2007)\citenamefont
  {Marenduzzo}, \citenamefont {Orlandini}, \citenamefont {Cates},\ and\
  \citenamefont {Yeomans}}]{marenduzzo2007steady}%
  \BibitemOpen
  \bibfield  {author} {\bibinfo {author} {\bibfnamefont {D.}~\bibnamefont
  {Marenduzzo}}, \bibinfo {author} {\bibfnamefont {E.}~\bibnamefont
  {Orlandini}}, \bibinfo {author} {\bibfnamefont {M.}~\bibnamefont {Cates}},\
  and\ \bibinfo {author} {\bibfnamefont {J.}~\bibnamefont {Yeomans}},\
  }\href@noop {} {\bibfield  {journal} {\bibinfo  {journal} {Physical Review
  E}\ }\textbf {\bibinfo {volume} {76}},\ \bibinfo {pages} {031921} (\bibinfo
  {year} {2007})}\BibitemShut {NoStop}%
\bibitem [{\citenamefont {Mart{\'\i}nez-Prat}\ \emph
  {et~al.}(2019)\citenamefont {Mart{\'\i}nez-Prat}, \citenamefont
  {Ign{\'e}s-Mullol}, \citenamefont {Casademunt},\ and\ \citenamefont
  {Sagu{\'e}s}}]{martinez2019selection}%
  \BibitemOpen
  \bibfield  {author} {\bibinfo {author} {\bibfnamefont {B.}~\bibnamefont
  {Mart{\'\i}nez-Prat}}, \bibinfo {author} {\bibfnamefont {J.}~\bibnamefont
  {Ign{\'e}s-Mullol}}, \bibinfo {author} {\bibfnamefont {J.}~\bibnamefont
  {Casademunt}},\ and\ \bibinfo {author} {\bibfnamefont {F.}~\bibnamefont
  {Sagu{\'e}s}},\ }\href@noop {} {\bibfield  {journal} {\bibinfo  {journal}
  {Nature Physics}\ }\textbf {\bibinfo {volume} {15}},\ \bibinfo {pages} {362}
  (\bibinfo {year} {2019})}\BibitemShut {NoStop}%
\bibitem [{\citenamefont {Mathijssen}\ \emph {et~al.}(2016)\citenamefont
  {Mathijssen}, \citenamefont {Doostmohammadi}, \citenamefont {Yeomans},\ and\
  \citenamefont {Shendruk}}]{mathijssen2016hydrodynamics}%
  \BibitemOpen
  \bibfield  {author} {\bibinfo {author} {\bibfnamefont {A.~J.}\ \bibnamefont
  {Mathijssen}}, \bibinfo {author} {\bibfnamefont {A.}~\bibnamefont
  {Doostmohammadi}}, \bibinfo {author} {\bibfnamefont {J.~M.}\ \bibnamefont
  {Yeomans}},\ and\ \bibinfo {author} {\bibfnamefont {T.~N.}\ \bibnamefont
  {Shendruk}},\ }\href@noop {} {\bibfield  {journal} {\bibinfo  {journal}
  {Journal of Fluid Mechanics}\ }\textbf {\bibinfo {volume} {806}},\ \bibinfo
  {pages} {35} (\bibinfo {year} {2016})}\BibitemShut {NoStop}%
\bibitem [{\citenamefont {Maitra}\ \emph {et~al.}(2018)\citenamefont {Maitra},
  \citenamefont {Srivastava}, \citenamefont {Marchetti}, \citenamefont
  {Lintuvuori}, \citenamefont {Ramaswamy},\ and\ \citenamefont
  {Lenz}}]{maitra2018nonequilibrium}%
  \BibitemOpen
  \bibfield  {author} {\bibinfo {author} {\bibfnamefont {A.}~\bibnamefont
  {Maitra}}, \bibinfo {author} {\bibfnamefont {P.}~\bibnamefont {Srivastava}},
  \bibinfo {author} {\bibfnamefont {M.~C.}\ \bibnamefont {Marchetti}}, \bibinfo
  {author} {\bibfnamefont {J.~S.}\ \bibnamefont {Lintuvuori}}, \bibinfo
  {author} {\bibfnamefont {S.}~\bibnamefont {Ramaswamy}},\ and\ \bibinfo
  {author} {\bibfnamefont {M.}~\bibnamefont {Lenz}},\ }\href@noop {} {\bibfield
   {journal} {\bibinfo  {journal} {Proceedings of the National Academy of
  Sciences}\ }\textbf {\bibinfo {volume} {115}},\ \bibinfo {pages} {6934}
  (\bibinfo {year} {2018})}\BibitemShut {NoStop}%
\bibitem [{\citenamefont {Sultan}\ \emph {et~al.}(2022)\citenamefont {Sultan},
  \citenamefont {Nejad},\ and\ \citenamefont
  {Doostmohammadi}}]{sultan2022quadrupolar}%
  \BibitemOpen
  \bibfield  {author} {\bibinfo {author} {\bibfnamefont {S.~A.}\ \bibnamefont
  {Sultan}}, \bibinfo {author} {\bibfnamefont {M.~R.}\ \bibnamefont {Nejad}},\
  and\ \bibinfo {author} {\bibfnamefont {A.}~\bibnamefont {Doostmohammadi}},\
  }\href@noop {} {\bibfield  {journal} {\bibinfo  {journal} {Soft Matter}\
  }\textbf {\bibinfo {volume} {18}},\ \bibinfo {pages} {4118} (\bibinfo {year}
  {2022})}\BibitemShut {NoStop}%
\bibitem [{\citenamefont {Lemma}\ \emph {et~al.}(2023)\citenamefont {Lemma},
  \citenamefont {Varghese}, \citenamefont {Ross}, \citenamefont {Thomson},
  \citenamefont {Baskaran},\ and\ \citenamefont {Dogic}}]{lemma2023spatio}%
  \BibitemOpen
  \bibfield  {author} {\bibinfo {author} {\bibfnamefont {L.~M.}\ \bibnamefont
  {Lemma}}, \bibinfo {author} {\bibfnamefont {M.}~\bibnamefont {Varghese}},
  \bibinfo {author} {\bibfnamefont {T.~D.}\ \bibnamefont {Ross}}, \bibinfo
  {author} {\bibfnamefont {M.}~\bibnamefont {Thomson}}, \bibinfo {author}
  {\bibfnamefont {A.}~\bibnamefont {Baskaran}},\ and\ \bibinfo {author}
  {\bibfnamefont {Z.}~\bibnamefont {Dogic}},\ }\href@noop {} {\bibfield
  {journal} {\bibinfo  {journal} {PNAS Nexus}\ }\textbf {\bibinfo {volume}
  {2}},\ \bibinfo {pages} {pgad130} (\bibinfo {year} {2023})}\BibitemShut
  {NoStop}%
\bibitem [{\citenamefont {Zarei}\ \emph {et~al.}(2023)\citenamefont {Zarei},
  \citenamefont {Berezney}, \citenamefont {Hensley}, \citenamefont {Lemma},
  \citenamefont {Senbil}, \citenamefont {Dogic},\ and\ \citenamefont
  {Fraden}}]{zarei2023light}%
  \BibitemOpen
  \bibfield  {author} {\bibinfo {author} {\bibfnamefont {Z.}~\bibnamefont
  {Zarei}}, \bibinfo {author} {\bibfnamefont {J.}~\bibnamefont {Berezney}},
  \bibinfo {author} {\bibfnamefont {A.}~\bibnamefont {Hensley}}, \bibinfo
  {author} {\bibfnamefont {L.}~\bibnamefont {Lemma}}, \bibinfo {author}
  {\bibfnamefont {N.}~\bibnamefont {Senbil}}, \bibinfo {author} {\bibfnamefont
  {Z.}~\bibnamefont {Dogic}},\ and\ \bibinfo {author} {\bibfnamefont
  {S.}~\bibnamefont {Fraden}},\ }\href@noop {} {\bibfield  {journal} {\bibinfo
  {journal} {Soft Mtter}\ }\textbf {\bibinfo {volume} {19}},\ \bibinfo {pages}
  {6691} (\bibinfo {year} {2023})}\BibitemShut {NoStop}%
\bibitem [{\citenamefont {Elosegui-Artola}(2021)}]{elosegui2021extracellular}%
  \BibitemOpen
  \bibfield  {author} {\bibinfo {author} {\bibfnamefont {A.}~\bibnamefont
  {Elosegui-Artola}},\ }\href@noop {} {\bibfield  {journal} {\bibinfo
  {journal} {Current Opinion in Cell Biology}\ }\textbf {\bibinfo {volume}
  {72}},\ \bibinfo {pages} {10} (\bibinfo {year} {2021})}\BibitemShut {NoStop}%
\bibitem [{\citenamefont {Flemming}\ \emph {et~al.}(2023)\citenamefont
  {Flemming}, \citenamefont {van Hullebusch}, \citenamefont {Neu},
  \citenamefont {Nielsen}, \citenamefont {Seviour}, \citenamefont {Stoodley},
  \citenamefont {Wingender},\ and\ \citenamefont
  {Wuertz}}]{flemming2023biofilm}%
  \BibitemOpen
  \bibfield  {author} {\bibinfo {author} {\bibfnamefont {H.-C.}\ \bibnamefont
  {Flemming}}, \bibinfo {author} {\bibfnamefont {E.~D.}\ \bibnamefont {van
  Hullebusch}}, \bibinfo {author} {\bibfnamefont {T.~R.}\ \bibnamefont {Neu}},
  \bibinfo {author} {\bibfnamefont {P.~H.}\ \bibnamefont {Nielsen}}, \bibinfo
  {author} {\bibfnamefont {T.}~\bibnamefont {Seviour}}, \bibinfo {author}
  {\bibfnamefont {P.}~\bibnamefont {Stoodley}}, \bibinfo {author}
  {\bibfnamefont {J.}~\bibnamefont {Wingender}},\ and\ \bibinfo {author}
  {\bibfnamefont {S.}~\bibnamefont {Wuertz}},\ }\href@noop {} {\bibfield
  {journal} {\bibinfo  {journal} {Nature Reviews Microbiology}\ }\textbf
  {\bibinfo {volume} {21}},\ \bibinfo {pages} {70} (\bibinfo {year}
  {2023})}\BibitemShut {NoStop}%
\bibitem [{\citenamefont {Cabral}\ and\ \citenamefont
  {Leedom}(1993)}]{cabral1993imaging}%
  \BibitemOpen
  \bibfield  {author} {\bibinfo {author} {\bibfnamefont {B.}~\bibnamefont
  {Cabral}}\ and\ \bibinfo {author} {\bibfnamefont {L.~C.}\ \bibnamefont
  {Leedom}},\ }in\ \href@noop {} {\emph {\bibinfo {booktitle} {Proceedings of
  the 20th annual conference on Computer graphics and interactive
  techniques}}}\ (\bibinfo {year} {1993})\ pp.\ \bibinfo {pages}
  {263--270}\BibitemShut {NoStop}%
\bibitem [{\citenamefont {van~der Walt}\ \emph {et~al.}(2014)\citenamefont
  {van~der Walt}, \citenamefont {Schönberger}, \citenamefont {Nunez-Iglesias},
  \citenamefont {Boulogne}, \citenamefont {Warner}, \citenamefont {Yager},
  \citenamefont {Gouillart}, \citenamefont {Yu},\ and\ \citenamefont {the
  scikit-image contributors}}]{vanderWalt2014}%
  \BibitemOpen
  \bibfield  {author} {\bibinfo {author} {\bibfnamefont {S.}~\bibnamefont
  {van~der Walt}}, \bibinfo {author} {\bibfnamefont {J.~L.}\ \bibnamefont
  {Schönberger}}, \bibinfo {author} {\bibfnamefont {J.}~\bibnamefont
  {Nunez-Iglesias}}, \bibinfo {author} {\bibfnamefont {F.}~\bibnamefont
  {Boulogne}}, \bibinfo {author} {\bibfnamefont {J.~D.}\ \bibnamefont
  {Warner}}, \bibinfo {author} {\bibfnamefont {N.}~\bibnamefont {Yager}},
  \bibinfo {author} {\bibfnamefont {E.}~\bibnamefont {Gouillart}}, \bibinfo
  {author} {\bibfnamefont {T.}~\bibnamefont {Yu}},\ and\ \bibinfo {author}
  {\bibnamefont {the scikit-image contributors}},\ }\href@noop {} {\bibfield
  {journal} {\bibinfo  {journal} {PeerJ}\ }\textbf {\bibinfo {volume} {2}},\
  \bibinfo {pages} {e453} (\bibinfo {year} {2014})}\BibitemShut {NoStop}%
\bibitem [{\citenamefont {Copenhagen}\ \emph {et~al.}(2021)\citenamefont
  {Copenhagen}, \citenamefont {Alert}, \citenamefont {Wingreen},\ and\
  \citenamefont {Shaevitz}}]{copenhagen2021topological}%
  \BibitemOpen
  \bibfield  {author} {\bibinfo {author} {\bibfnamefont {K.}~\bibnamefont
  {Copenhagen}}, \bibinfo {author} {\bibfnamefont {R.}~\bibnamefont {Alert}},
  \bibinfo {author} {\bibfnamefont {N.~S.}\ \bibnamefont {Wingreen}},\ and\
  \bibinfo {author} {\bibfnamefont {J.~W.}\ \bibnamefont {Shaevitz}},\
  }\href@noop {} {\bibfield  {journal} {\bibinfo  {journal} {Nature Physics}\
  }\textbf {\bibinfo {volume} {17}},\ \bibinfo {pages} {211} (\bibinfo {year}
  {2021})}\BibitemShut {NoStop}%
\bibitem [{\citenamefont {Kawaguchi}\ \emph {et~al.}(2017)\citenamefont
  {Kawaguchi}, \citenamefont {Kageyama},\ and\ \citenamefont
  {Sano}}]{kawaguchi2017topological}%
  \BibitemOpen
  \bibfield  {author} {\bibinfo {author} {\bibfnamefont {K.}~\bibnamefont
  {Kawaguchi}}, \bibinfo {author} {\bibfnamefont {R.}~\bibnamefont
  {Kageyama}},\ and\ \bibinfo {author} {\bibfnamefont {M.}~\bibnamefont
  {Sano}},\ }\href@noop {} {\bibfield  {journal} {\bibinfo  {journal} {Nature}\
  }\textbf {\bibinfo {volume} {545}},\ \bibinfo {pages} {327} (\bibinfo {year}
  {2017})}\BibitemShut {NoStop}%
\end{thebibliography}%
\bibliographystyle{apsrev4-2}
\end{document}